\documentclass[preprint2]{aastex}
\usepackage{graphicx,amssymb}

\shorttitle{Molecular outflows in the young open cluster IC\,348}
\shortauthors{Eisl\"offel et al.}

\begin{document}
  
\title{Molecular outflows in the young open cluster IC\,348\thanks{Based 
on observations taken at the German-Spanish Astronomical Centre, Calar 
Alto, operated by the Max-Planck-Institute for Astronomy, Heidelberg, 
jointly with the Spanish National Commission for Astronomy and on 
observations with ISO, an ESA project with instruments funded by ESA 
Member States (especially the PI countries: France, Germany, the 
Netherlands and the United Kingdom) and with the participation 
of ISAS and NASA.}}

\author{Jochen Eisl\"offel, Dirk Froebrich}
\affil{Th\"uringer Landessternwarte Tautenburg, Sternwarte 5, 
       D-07778 Tautenburg, Germany}

\author{Thomas Stanke}
\affil{Max-Planck-Institut f\"ur Radioastronomie, Auf dem H\"ugel 69,
       D-53121 Bonn, Germany}

\author{Mark J. McCaughrean}
\affil{Astrophysikalisches Institut Potsdam, An der Sternwarte 16,
       D-14482 Potsdam, Germany}

\begin{abstract}
We present a wide-field survey of the young open cluster IC\,348 for
molecular H$_2$ outflows. Outflow activity is only found at its 
south-western limit, where a new subcluster of embedded sources
is in an early phase of its formation. If the IC\,348 cluster had been
built up by such subclusters forming at different times, this could
explain the large age-spread that \citet{h98} found for the IC\,348
member stars. In addition to several compact groups of H$_2$ knots,
our survey reveals a large north-south oriented outflow, and we
identify the newly discovered far-infrared and mm-object IC\,348\,MMS
as its source. New deep images in the 1--0 S(1) line of molecular
hydrogen trace the HH\,211 jet and counterjet as highly-collimated
chains of knots, resembling the interferometric CO and SiO jets. This
jet system appears rotated counter-clockwise by about 3$\degr$ with
respect to the prominent H$_2$ bow shocks. Furthermore, we resolve
HH\,211-mm as a double point-like source in the mm-continuum.
\end{abstract}

\keywords{Shock waves -- ISM: jets and outflows --  ISM: kinematics 
and dynamics -- ISM: molecules -- stars: mass-loss}


\section{Introduction}

The early stages of star formation are acompanied by powerful 
bipolar outflows. The investigation of such outflows helps us to 
better understand the star formation process as a whole. On the other
hand, the flows are also useful as pointers to their sources, which
in the earliest phases of their formation are deeply embedded in their
parental clouds and not visible at optical or near-infrared
wavelengths. Once the flows start breaking out of these dense
cores into regions of lower extinction, the shock excited gas 
produced on their way through the ambient medium becomes
observable in the near-infrared. An excellent tracer for these flows
then is the emission of molecular hydrogen in its 1--0 S(1) line at
2.122\,$\mu$m.

IC\,348 is a young open cluster in the Perseus dark cloud. Several
hundred cluster members have been found in the optical \citep{h98}, 
in the near-infrared \citep{ll95, lrll98, ntc00} and in X-rays 
\citep{pzh96, pz01}. \citet{h98} investigated the age of the cluster
and found a mean age of 1.3\,Myr, with a surprisingly large spread
around this value of 0.7 to 12\,Myr. This large spread is quite
unusual for such a young cluster, and it may point to an unusual
formation history. Since Herbig derived the ages only for optically
visible stars, his lower limit may also imply that star formation is
still going on in this cluster: There may be even younger objects, 
which cannot be detected in the optical because they are still too  
deeply embedded.

Indeed, \citet{ssc74} made an IR-map of IC\,348, and found a bright 
IR source (IC\,348-IR) in a small red reflection nebula, which was 
later studied at high spatial resolution by \citet{bcmnr95}. In its 
neighbourhood \citet{mrz94} discovered a highly-collimated molecular
outflow, HH\,211, in the south-western part of the cluster. In this
region \citet{bgk87} had already found several large clumps in ammonia
and CO. The HH\,211 flow has subsequently been studied in detail in
high-resolution interferometric observations in CO by \citet{gg99} and
in SiO by \citet{cr01}. It source is suspected to be a Class\,0
object, a protostar in its earliest stage of formation going through 
its main accretion phase. 

On the K$'$ mosaic of \citet{mrz94} a chain of knots was seen north of
HH\,211, that could be part of another outflow. This prompted us to
carry out a survey in molecular hydrogen of the entire IC\,348 cluster
to search for outflows and identify their sources, as a possibly so
far unknown very young population in this cluster.

In the following we first describe our observations and data reduction
in Sect.\,\ref{obs}. We then present our new deep imaging of the 
HH\,211 flow in Sect.\ref{hh211}, and continue with the newly found
H$_2$ knots and flows in Sect.\,\ref{newflows}. Sect.\,\ref{sources}
is devoted to a discussion of the possible outflow sources and the
recent star formation in IC\,348.


\section{Observations and Data Reduction}
\label{obs}

\subsection{Near-infrared data}

Our near-infrared data were taken during several observing runs at the 
3.5-m telescope on Calar Alto. In November 1995 we observed the region 
of the HH\,211 outflow with the MAGIC infrared camera \citep{hbbhmmw93}. 
The high resolution optics yielded a scale of 0\farcs32 
per pixel. Narrow-band filters centered on the 1--0\,S(1) line of 
molecular hydrogen at 2.122\,$\mu$m and on the nearby continuum at 
2.140\,$\mu$m were used. We observed a small mosaic with a per pixel 
integration time of 1740\,s. The seeing was about 0\farcs9. These data 
were not flux calibrated, because the weather conditions were not 
photometric.  

In September 1997 the whole IC\,348 cluster was observed under photometric 
weather conditions. This time, we used MAGIC with its wide field optic with 
a scale of 0\farcs81 per pixel. Again the narrow-band filter centered
at 2.122\,$\mu$m with a FWHM of 0.019\,$\mu$m was used. The 
per-pixel integration time was 180\,s, and the seeing was about 1\farcs2.
Deeper mosaics in the H$_2$ 1-0 S(1) line and in the continuum were taken 
of the region of the newly discovered outflows. They have integration times 
of 620\,s and 520\,s per pixel, and have been photometrically calibrated 
using the standards HD\,18881, HD\,203856, HD\,3029 and HD\,40335. In 
September 1997 a 13$'$ $\times$ 13$'$ area around HH\,211 was also observed 
with the Omega Prime camera \citep{bmbts98}, at lower sensitivity, but 
covering an area extending further south than the MAGIC data.

In December 2000 we used again the Omega Prime camera with a scale of 
0\farcs396 and a field of view of 6\farcm8 $\times$ 6\farcm8. The 
weather conditions were photometric and the seeing about 1\farcs5. 
Again we used a H$_2$ 1--0\,S(1) narrow-band filter at 2.122\,$\mu$m, 
with an integration time per pixel of 810\,s. For flux calibration we 
observed the standard star FS13.

Flatfielding of the images was done using lamp on/off flats. For sky
subtraction and co-addition we used the DIMSUM package in IRAF. The 
single images were co-centred onto a large reference frame using all 
detectable stars in the field, to ensure a high astrometric accuracy. 
Standard star images were reduced in the same way. Photometric errors 
are on the order of 0.05\,mag. We used HST guide stars to determine 
the astrometric solution and reached a positional accuracy of about 
0\farcs5. Emission features were identified by comparing the 
images taken in  H$_2$ and in the continuum in all cases. 

\begin{figure*}[t]
\centering
\resizebox{16cm}{!}{\includegraphics[angle=-90,bb=150 80 520 790]{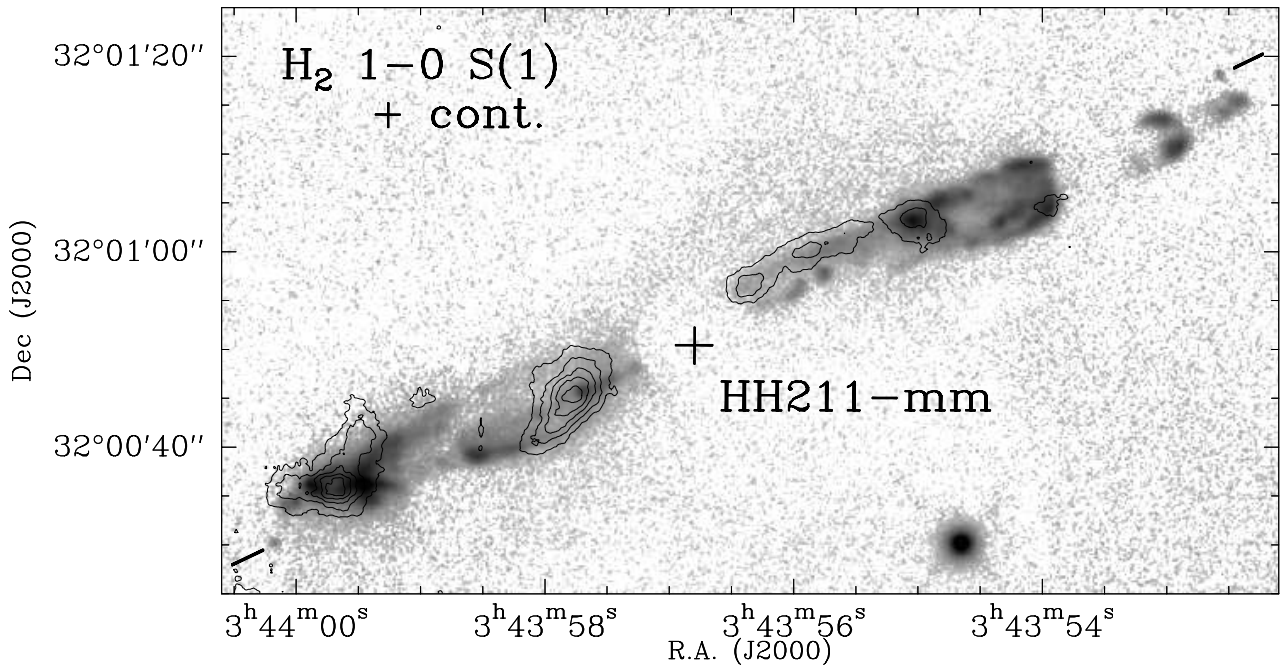}}
\caption{Grey-scale plot of the HH\,211 outflow in the 1--0\,S(1) 
line of molecular hydrogen at 2.122\,$\mu$m, with logarithmic
intensities. The contours show the continuum at 2.140\,$\mu$m. 
A cross marks the position of the outflow source HH\,211-mm. The
orientation of the highly-collimated knotty jet and counterjet seen in
H$_2$ is marked by solid lines on both ends. This jet system is rotated
counter-clockwise by about 3$\degr$ with respect to the system of the
prominent H$_2$ bows. The two point-like features next to the solid lines 
are likely background stars.}
\label{hh211_deep}
\end{figure*}

\subsection{Far-infrared and mm data}

Small maps of the HH\,211 region were taken with the PHOT instrument 
onboard the ISO satellite in its PHT22 mode and the C200 detector by 
stepping the instrument by half a detector pixel. This way, 7$\times$3 
mosaics with a pixel size of 45$''$$\times$90$''$, and a total field 
of view of 315$''$$\times$270$''$ were
obtained. The data were reduced with the ISOPHOT Interactive Analysis 
(PIA V9.1) software. Fluxes of the sources were measured by attributing 
pixels manually to ``object'' or ``background'', then summing up the 
object flux and subtracting the background.

Our mm-continuum observations were obtained in March 1999 at the IRAM
30-m antenna using the 37-channel bolometer array MAMBO \citep{kgg98} 
at a wavelength of 1.2\,mm. An area of about 36\,$\Box$\arcmin\ was
covered. The data were reduced with MOPSI (software package
developed by R. Zylka, pers. comm.) following standard procedures. 
After an initial despiking and low-order baseline subtraction the data 
were corrected for atmospheric extinction and flux calibrated. Then 
correlated sky brightness variations ("sky-noise") were removed in an 
iterative manner, using the result of previous runs as an input 
source model to improve the performance of the sky-noise removal. 
After further despiking and residual baseline subtraction the data 
were restored from dual beam to single beam and finally converted 
into a map.

\section{The HH\,211 outflow}
\label{hh211}

In Fig.\,\ref{hh211_deep} we present a new image of HH\,211 in 
the 1--0\,S(1) line of molecular hydrogen at 2.122\,$\mu$m. This 
deeper image reveals some important details, which were not 
visible on the original data by \citet{mrz94}. Most 
interestingly, we clearly see highly-collimated chains of knots in the
jet and the counterjet. They are tracing the high-velocity molecular
flows, which are seen in the interferometric observations in CO by
\citet{gg99} and in SiO by \citet{cr01}. This chain of knots is 
running roughly through the middle of the rim-brightened cavity-like 
structure. 
A line drawn through the knots marking the high-velocity jet
(indicated by the solid lines at both ends of the jet in
Fig.\,\ref{hh211_deep}) does not, however, exactly lead through the
tips of the prominent H$_2$ bows. Apparently this jet system is
rotated about 3$\degr$ counter-clockwise with respect to the
broader outflow lobes. This misalignment, which is also seen in some
other jets, may be indicative of precession in the HH\,211
flow. Indeed, we will show in Sect.\,\ref{sources} that the HH\,211
source may be a binary. 

Comparison with continuum images also shows that part of the
rim-brightened cavity -- especially the part close to the source and
almost parallel to the western jet -- is not H$_2$, but instead
continuum emission (see the contour plot in Fig.\,\ref{hh211_deep}, 
which shows the continuum at 2.140\,$\mu$m). Here we are probably 
seeing scattered light from the source that escapes from the dense 
shielding envelope through the jet channels, thus opening up the 
possibility of indirectly obtaining a spectrum of HH\,211-mm.

\begin{figure}[th]
\resizebox{7.5cm}{!}{\includegraphics[angle=-90,bb=55 115 481 380]{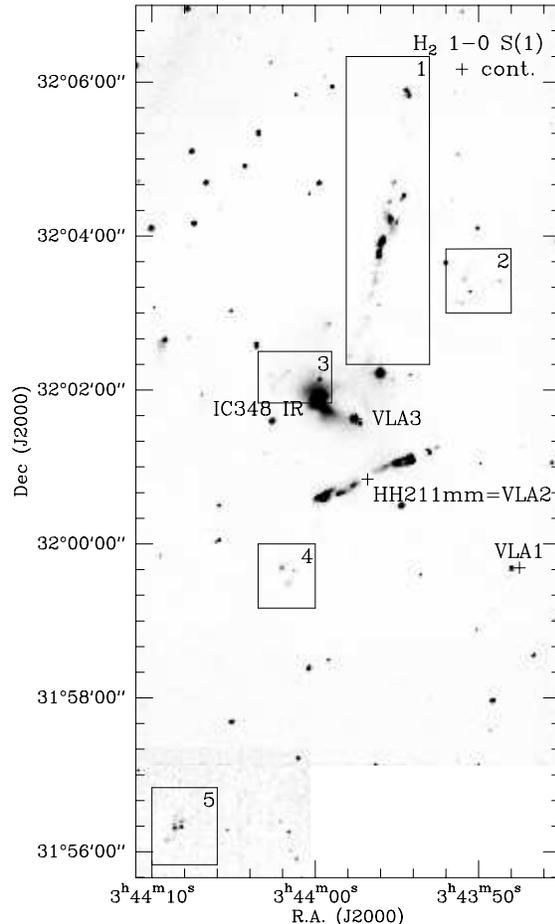}}
\caption{HH211 and the newly discovered H$_2$ emission knots at 
2.122\,$\mu$m. Boxes mark the regions which are displayed in detail as 
contour plots. The + signs mark the 3.5\,cm sources detected by \citet{arc01}.}
\label{hh211_2122_grayscale}
\end{figure}

\section{New outflows and H$_2$ emission features in IC\,348}
\label{newflows}

Our full H$_2$ survey covers a 390\,$\Box$\arcmin\ region centered on 
the young cluster 
IC\,348 (between R.A.(J2000) $3^h43^m42\fs3$ and $3^h44^m56\fs6$, 
and Dec(J2000) $+31 \degr 56'00''$ and $+32 \degr 19'52''$) 
in the 1--0\,S(1) line of molecular hydrogen at 2.122\,$\mu$m. 
Outflow activity was only detected towards the south-western limit of 
the cluster, in the region of the well-known HH\,211 outflow (see
Fig.\,\ref{hh211_2122_grayscale}). This region coincides with the area 
in which \citet{ssc74} had found the embedded infrared source IC\,348-IR 
and \citet{bgk87} found several clumps in ammonia and CO. These 
signs show that star formation is still going on in this part of the 
IC\,348 cluster. 

\begin{figure}[th]
\vspace{-5mm}
\centering\resizebox{5.5cm}{!}{\includegraphics[angle=-90,bb=55 50 530 260]{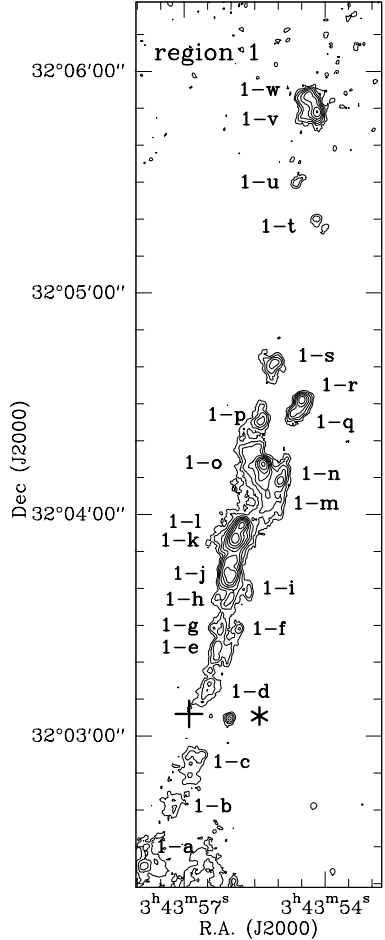}}
\caption{Contour plot of region 1 at 2.122\,$\mu$m. The contours start at 
the 1.5\,$\times$\,10$^{-19}$\,Wm$^{-2}$arcsec$^{-2}$ and then increase by a 
factor of $\sqrt 3$. The ``+'' sign marks the position of
IC\,348\,MMS, while the ``$\ast$'' means the object is not a H$_2$ 
emission feature but a star.}
\label{hh211_2122_cont1}
\end{figure}

\begin{figure}[th]
\centering
\resizebox{7.0cm}{!}{\includegraphics[angle=-90,bb=50 55 510 650]{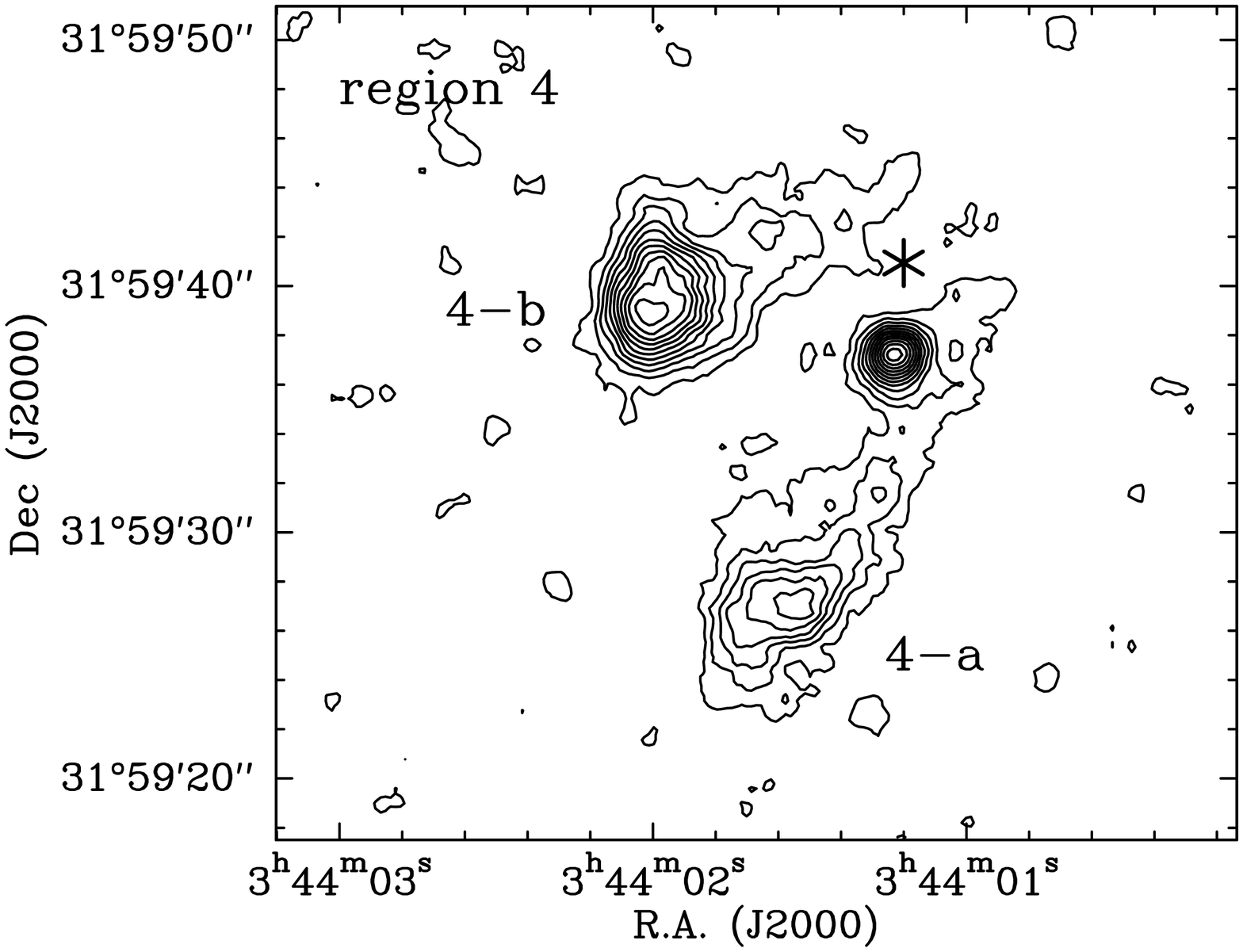}} \\
\vspace*{5mm}
\resizebox{7.0cm}{!}{\includegraphics[angle=-90,bb=50 70 510 670]{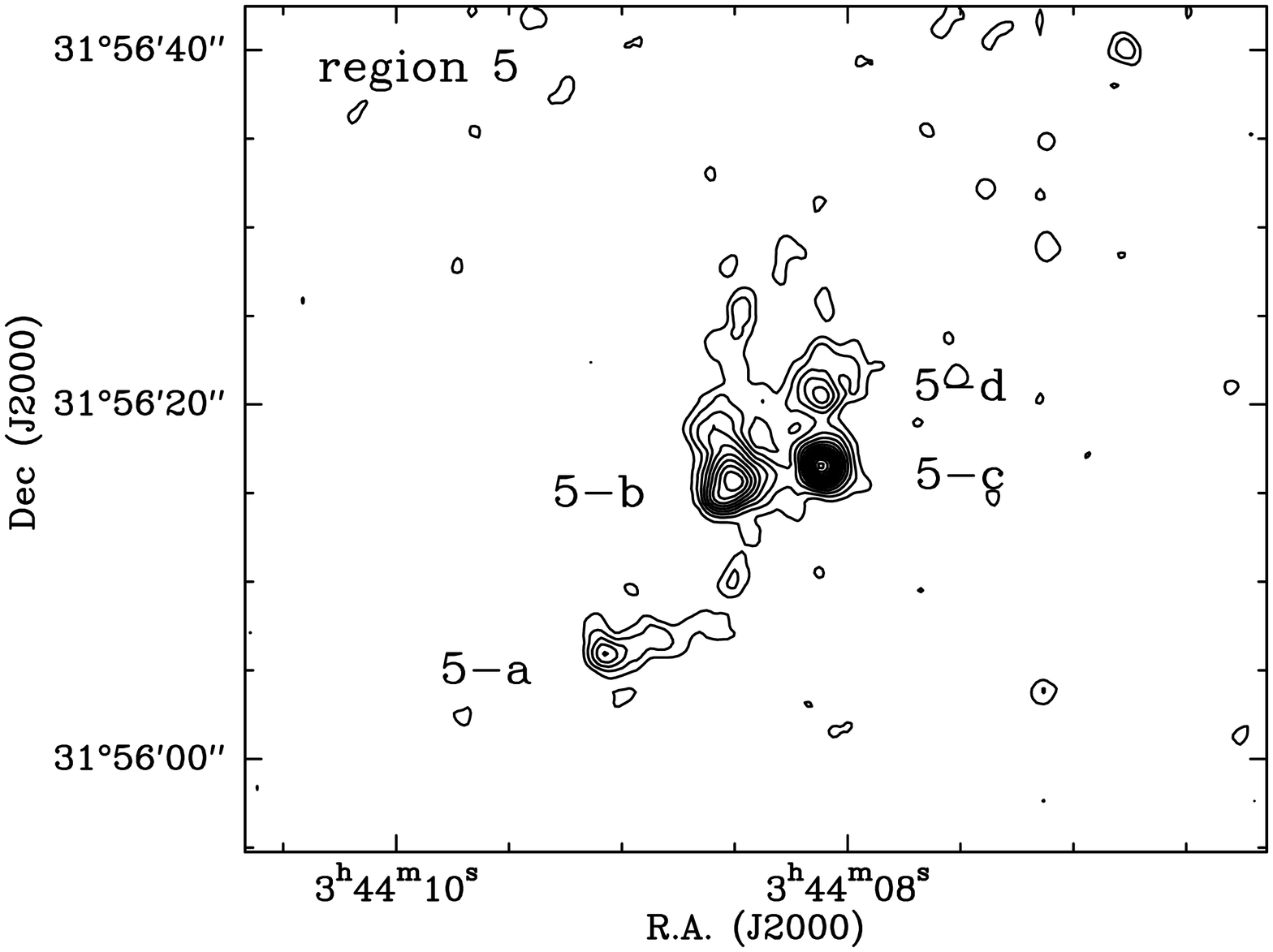}}
\caption{Contour plot of region 4 (upper panel) and region 5 (lower panel) 
at 2.122\,$\mu$m. The contours start at 
1.5\,$\times$\,10$^{-19}$\,Wm$^{-2}$arcsec$^{-2}$ in both panels, but 
increase by a factor of $\sqrt 3$ in the upper panel and linearly by 
7.5\,$\times$\,10$^{-20}$\,Wm$^{-2}$arcsec$^{-2}$. The ``$\ast$'' 
marks a star.}
\label{hh211_2122_cont4}
\end{figure}

\begin{figure}[th]
\resizebox{7.5cm}{!}{\includegraphics[angle=-90,bb=50 55 510 640]{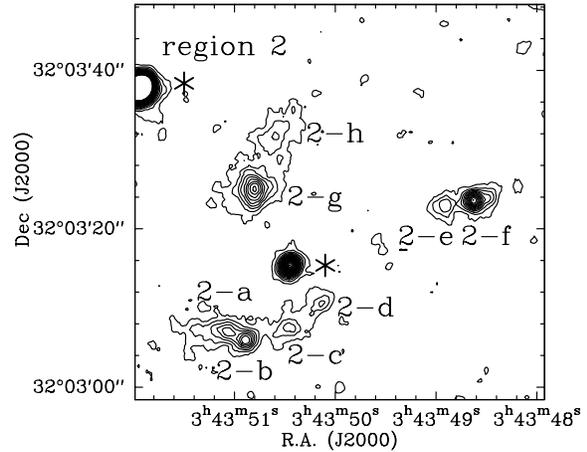}}
\caption{Contour plot of region 2 at 2.122\,$\mu$m. The contours start at 
the 1\,$\times$\,10$^{-19}$\,Wm$^{-2}$arcsec$^{-2}$ level and increase 
linearly in steps of 7.5\,$\times$\,10$^{-20}$\,Wm$^{-2}$arcsec$^{-2}$. 
The ``$\ast$'' sign means the object is a star, not a H$_2$ emission 
feature.}
\label{hh211_2122_cont2}
\end{figure}

\begin{figure}[ht]
\resizebox{7.5cm}{!}{\includegraphics[angle=-90,bb=50 55 510 755]{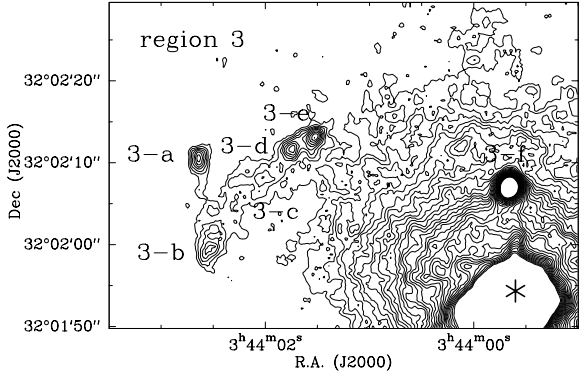}}
\caption{Contour plot of region 3 at 2.122\,$\mu$m. The contours start at 
the 1.8\,$\times$\,10$^{-19}$\,Wm$^{-2}$arcsec$^{-2}$ level and increase 
linearly in steps of 5\,$\times$\,10$^{-20}$\,Wm$^{-2}$arcsec$^{-2}$. 
The ``$\ast$'' marks IC348\,IR .} 
\label{hh211_2122_cont3}
\end{figure}

Apart from HH\,211, we find H$_2$ emission features spreading over
five regions near HH\,211, which are outlined in
Fig.\,\ref{hh211_2122_grayscale}. These regions labelled 1 to 5 are
shown enlarged as contour plots in Figs.\,\ref{hh211_2122_cont1} to 
\ref{hh211_2122_cont3}. In these plots, the more prominent H$_2$ 
emission features have been labelled, and their positions and fluxes 
are given in Table\,\ref{positions}. Also marked are possible outflow
sources in the field: the ``+'' signs show the positions of 3.5\,cm 
sources detected by \citet{arc01}. 

The most prominent of these new H$_2$ features is a large, roughly
north-south oriented outflow. Its brightest parts could already be
seen on the K$'$ image of \citet{mrz94}, where it was referred to 
as the ``chain'' of knots. Its northern part is shown in 
Fig.\,\ref{hh211_2122_cont1}. It consists of a strand of knots, almost
4\,arcmin long, which splits up in the region between knots 1-l and
1-s. Here, we are probably seeing condensations that are part of a
large bow shock. After a gap of about 30$''$, a few more H$_2$ knots
are seen, which end in the bow-shaped knots 1-v/w. The flow shows a
clearly visible bending, of about 10$\degr$ towards the east, over its
length of almost 4$'$. Such bending is also observed in several other
jets and outflows (see, e.g. \citet{e00}). A variety of possible
explanations have been given for such bending by \citet{fz98},
\citet{tepn99}, and \citet{e00}: Lorentz forces on a magnetic jet,
motion of the source in a binary system, warping of the disk or
dynamical pressure of the external medium. Here, we do not yet have 
enough information about the kinematics of the flow's knots available
that would permit us to distinguish between these possibilities. 

\begin{table}[thp]
\caption{\label{positions}Positions and fluxes of newly found 
H$_2$ knots. Positional errors are on the order of 0\farcs5. Photometric
errors range from 5\% for the brighter and up to 20\% for the fainter
knots.}
\begin{center}
\renewcommand{\baselinestretch}{1.0}
\renewcommand{\tabcolsep}{4pt}
{\scriptsize
\begin{tabular}{lccrr}
Object & $\alpha$(J2000) & $\delta$(J2000) & Flux$^*$ & Surface$^{**}$ \\
& & & & Brightness \\ \hline
1-a & 03:43:57.8 & +32:02:25 & 14.3 & 7.2 \\
1-b & 03:43:57.2 & +32:02:43 & 10.2 & 3.1 \\
1-c & 03:43:56.6 & +32:02:56 & 18.3 & 5.0 \\
1-d & 03:43:56.5 & +32:03:14 & 17.3 & 5.4 \\
1-e & 03:43:56.3 & +32:03:25 & 21.9 & 7.0 \\
1-f & 03:43:55.9 & +32:03:30 &  4.7 & 5.5 \\
1-g & 03:43:56.3 & +32:03:30 & 11.8 & 6.7 \\
1-h & 03:43:56.1 & +32:03:38 & 13.4 & 7.1 \\
1-i & 03:43:55.7 & +32:03:40 &  6.0 & 4.3 \\
1-j & 03:43:56.1 & +32:03:44 & 91.0 & 41.0 \\
1-k & 03:43:56.0 & +32:03:54 & & 120.6 \\
1-l & 03:43:55.8 & +32:03:58 & \raisebox{1.5ex}[-1.5ex]{$\big\}$199.4} & 101.5 \\
1-m & 03:43:55.3 & +32:04:05 & 30.9 & 9.3 \\
1-n & 03:43:55.0 & +32:04:10 & 23.8 & 24.2 \\
1-o & 03:43:55.4 & +32:04:15 & 114.0 & 92.5 \\
1-p & 03:43:55.4 & +32:04:27 & 17.8 & 11.6 \\
1-q & 03:43:54.8 & +32:04:28 & & 12.9 \\
1-r & 03:43:54.6 & +32:04:32 & \raisebox{1.5ex}[-1.5ex]{$\big\}$~~39.3} & 40.3 \\
1-s & 03:43:55.2 & +32:04:42 & 19.8 & 12.2 \\
1-t & 03:43:54.3 & +32:05:21 & 6.5 & 4.2 \\
1-u & 03:43:54.7 & +32:05:31 & 6.6 & 3.9 \\
1-v & 03:43:54.3 & +32:05:50 & & 47.7 \\
1-w & 03:43:54.5 & +32:05:54 & \raisebox{1.5ex}[-1.5ex]{$\big\}$~~71.9} & 22.5 \\
2-a & 03:43:51.1 & +32:03:08 & & 4.8 \\
2-b & 03:43:50.9 & +32:03:08 & \raisebox{1.5ex}[-1.5ex]{$\big\}$~~10.2} & 7.0 \\
2-c & 03:43:50.5 & +32:03:09 & 4.0 & 3.2 \\
2-d & 03:43:50.1 & +32:03:12 & 2.5 & 3.2 \\
2-e & 03:43:48.9 & +32:03:25 & 2.8 & 3.8 \\
2-f & 03:43:48.6 & +32:03:25 & 6.4 & 12.1 \\
2-g & 03:43:50.8 & +32:03:26 & 11.8 & 7.8 \\
2-h & 03:43:50.7 & +32:03:33 & 6.9 & 3.1 \\
3-a & 03:44:02.6 & +32:02:11 & 1.1 & 5.4 \\
3-b & 03:44:02.5 & +32:02:00 & 3.5 & 4.4 \\
3-c & 03:44:02.0 & +32:02:08 & 2.8 & 3.6 \\
3-d & 03:44:01.7 & +32:02:12 & & 5.5 \\
3-e & 03:44:01.5 & +32:02:14 & \raisebox{1.5ex}[-1.5ex]{$\big\}$~~~~4.8} & 7.1 \\
3-f & 03:43:59.6 & +32:02:08 & 11.5 & 30.4 \\
4-a & 03:44:01.4 & +31:59:27 & 14.8 & 5.3 \\
4-b & 03:44:01.8 & +31:59:40 & 20.8 & 9.2 \\
5-a & 03:44:08.7 & +31:56:05 & 5.2 & 4.1 \\
5-b & 03:44:08.2 & +31:56:15 & 10.8 & 7.6 \\
5-c & 03:44:07.8 & +31:56:16 & 6.4 & 10.2 \\
5-d & 03:44:07.8 & +31:56:20 & 5.2 & 4.3 \\
\end{tabular}}
\end{center}
\begin{list}{}{}
\item[$^{\mathrm{*}}$]Fluxes are in $10^{-18}$\,W\,m$^{-2}$ summed over 
the whole knot.
\item[$^{\mathrm{**}}$]Surface brightness of H$_2$ 1--0\,S(1) averaged 
over a 1\,\arcsec\ diameter are in $10^{-19}$\,W\,m$^{-2}$\,arcsec$^{-2}$. 
\end{list}
\end{table}

South of HH\,211, in regions 4 and 5 (see
Fig.\,\ref{hh211_2122_cont4}), we find two more compact groups of
H$_2$ knots, which we interpret as southward pointing bow shocks.
These bows have positions and orientations which suggest that they
belong to the southern lobe of the large north-south outflow. They 
are not aligned on a straight line with the northern flow, but instead
also show a clear bending of the flow towards the east in the southern
lobe.
The total length of this flow would then be 10$'$ or 0.87\,pc at a
distance of 300\,pc. We note, that in projection this flow passes very
close by the south-eastern bow of HH\,211. Although we do not detect
H$_2$ emission from it in this region, observations of HH\,211 in
other tracers may do so: In their interferometric study of CO in the
HH\,211 flow, \citet{gg99} found CO emission at 8.2\,km\,s$^{-1}$ in
this region, which was hard to explain if it came from HH\,211
itself. We suggest that this emission may not belong to the HH\,211
outflow, but instead comes from the large north-south flow.

In addition to the large north-south flow, we find two more regions of
H$_2$ emission north of HH\,211. These regions 2 and 3 are shown in
detail in Figs.\,\ref{hh211_2122_cont2} and
\ref{hh211_2122_cont3}. Both consist of compact groups of knots, which
again resemble incomplete bow shocks, that may be pointing south-east
(region 2) and east (region 3). Because of their orientations, we
speculate that these objects do not belong to the same outflow. We
note, however, that the bows in regions 2 and 5 are on a line that
passes very close by the HH\,211 source. Therefore, it cannot be
excluded that they form a flow emanating from the HH\,211 source as
well (which then should be a binary).

For the new north-south outflow we measure a total flux in the 1--0\,S(1)
line of H$_2$ of 6.8\,$\times$\,10$^{-16}$\,Wm$^{-2}$. This converts 
to a luminosity of about 1.9\,$\times$\,10$^{-3}$\,L$_{\odot}$ in this
line. From this value we can roughly infer the intrinsic total H$_2$ 
luminosity of this flow: In order to estimate the flux in all H$_2$ 
lines, we assume about 10\% to be emitted in the 1--0 S(1) line alone 
(which is the case, e.g., for C-shocks of 30\,--\,40\,kms$^{-1}$ shock 
speed, typical for such flows \citep{s91}. With an average
extinction of A$_K \approx$\,2\,mag for this region of the young 
IC\,348 cluster (see \citet{mrz94}), we find a total instrinsic H$_2$ 
luminosity for the flux in region 1 
of 0.1\,L$_{\odot}$. For the other regions we measure the following 
fluxes in the 1--0 S(1) line of H$_2$  
3.3\,$\times$\,10$^{-17}$ Wm$^{-2}$ (region 2),  
1.6\,$\times$\,10$^{-17}$ Wm$^{-2}$ (region 3),
3.6\,$\times$\,10$^{-17}$ Wm$^{-2}$ (region 4) and 
3.8\,$\times$\,10$^{-17}$\,W\,m$^{-2}$ (region 5), 
and estimate the intrinsic total H$_2$ luminosity in the same way as 
4.7, 2.3, 5.1, 5.4\,$\times$\,10$^{-3}$\,L$_{\odot}$, respectively.
Due to overlap effects the total flux for a region may slightly differ 
from a sum of the respective knots as given in Table \ref{positions}.

We also searched for morphological and photometric variability in the
H$_2$ knots. Such variability should be expected due to the short
cooling times in H$_2$ of order one year for the shock excited gas in 
the outflows. Nonetheless, we find at most very small or no changes at
all in the brightness of the knots over the five year timespan that
our H$_2$ observations cover.

Since from morphology alone it is difficult to assign the newly found
molecular emission features to individual outflows, we tried to
determine preliminary proper motions for the knots. Due to the coarse
pixel scale, especially of our earliest observations, the small number
of available epochs and the usable epoch difference of only 3.2 years,
we currently reach a measurement accuracy of only
300\,km\,s$^{-1}$. None of our measured knots is found to move that
fast. This is not unexpected, since in most flows for which precise
proper motions could be determined, values of around 200\,km\,s$^{-1}$
were found (see, e.g., \citet{em92}, \citet{em94}, \citet{mdres98}).

\section{Outflow sources and star formation in IC\,348}
\label{sources}

In a survey of the region at 3.5\,cm, \citet{arc01} found three VLA
sources, where VLA2 is identical with HH\,211-mm, while another
variable source VLA3 was found very close to the star
Cl*\,IC348\,LRL49. They suggested that this star may be the driving
source of the north-south outflow.

\begin{figure}[t]
\centering
\resizebox{7.5cm}{!}{\includegraphics[angle=0]{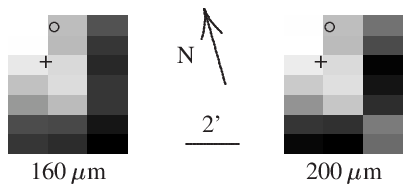}}
\caption{ISOPHOT maps at 160\,$\mu$m and 200\,$\mu$m centered on
HH\,211-mm. Also marked are IC\,348\,IR (plus) and IC\,348\,MMS (circle). 
Bright sources are white and sky background is black.}
\label{hh211_iso}
\end{figure}

\begin{figure}[t]
\centering
\resizebox{7.5cm}{!}{\includegraphics[angle=-90,bb=40 55 530 580]{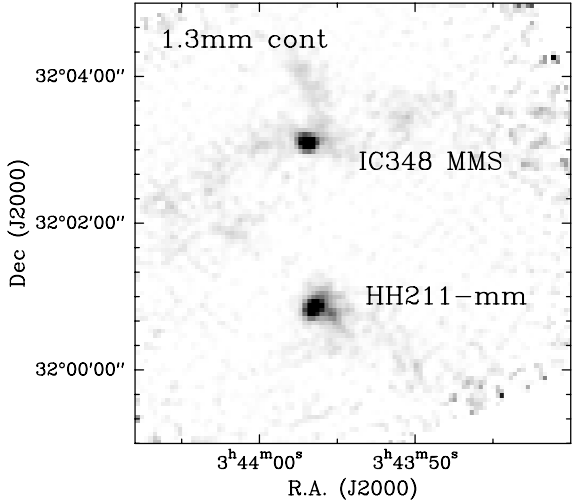}}
\caption{MAMBO 1.3\,mm map showing the HH\,211-mm (south) and the newly
discovered IC\,348\,MMS source (north).}
\label{hh211_mm}
\end{figure}

In order to search for possible exciting sources of the new molecular
outflow features, we obtained maps with ISOPHOT at 160\,$\mu$m and
200\,$\mu$m, as well as mm-continuum data with the MAMBO bolometer
array at 1.2\,mm. Our ISOPHOT maps (Fig.\,\ref{hh211_iso}) were
centred on HH\,211-mm. Strong emission from this source was detected
at both wavelengths. It is, however, affected by emission from other
point or extended sources to the north and north-east, which are not
resolved at the resolution of ISO. Our mm-map (Fig.\,\ref{hh211_mm}),
by contrast, clearly resolves point sources and extended emission.

The most prominent source in the field is HH\,211-mm. There appears to
be a previously unknown, fainter point-like source to the south-west. 
This second source is also indicated as a small elongation of HH\,211-mm 
in the submillimetre maps of \citet{cr00}, especially at 450\,$\mu$m and 
750\,$\mu$m. It was not, however, detected by \citet{arc01} at 
7\,mm and 3.5\,cm. The available data do not yet allow us to discern 
whether this object is a secondary to HH\,211-mm or only a dust 
condensation at the edge of the outflow cavity.

Both point
sources are embedded in a halo of diffuse emission, which extends
further to the south-west to the edge of our map. This ``dust ridge''
coincides with the H$^{13}$CO$^+$ filament seen by \citet{gg99}. In a
30$''$ aperture centred on HH\,211-mm (at a position of
$3^h43^m56\fs5$, $+32 \degr 00'51''$ (J2000)), we measure a total flux
of 1.88\,Jy, of which 0.89\,Jy come from the HH\,211-mm point source
itself, while the rest has to be attributed to the diffuse halo and
the ``companion''. 

About 2\,arcmin north of HH\,211-mm we find a hitherto unknown
mm-source, which we call IC\,348\,MMS. This source is also detected
in the ISOPHOT maps, where its position (at $3^h43^m56\fs9$, 
$+32 \degr 03'06''$ (J2000)) has been marked with a circle. It is
situated on a saddle point on an emission ridge between two of the
NH$_3$ (1,1) cores observed by \citet{bgk87}. This ridge and the
cores are also seen as faint dust filaments on our continuum map. 
We measure a flux of 1.58\,Jy in a 30$''$ aperture for this source. 

Since IC\,348\,MMS is situated in the gap between the H$_2$ knots 1-c and 1-d
(see Fig.\,\ref{hh211_2122_cont1}), we suggest it to be the driving
source of the long north-south molecular outflow. Measurements of the
radial velocities of the H$_2$ knots to its north and south should
confirm that this object and not VLA3 drives the long outflow.
The fact that this source is bright at far-infrared and
mm-wavelengths, but is not seen at all in the near-infrared makes it
likely that it is in the Class\,0 stage of its evolution, similar 
to HH\,211-mm: 
in our ISOPHOT maps we measure a flux of 55.3 and 48.2\,Jy at 160\,$\mu$m 
and 200\,$\mu$m, respectively, for IC\,348\,MMS (see \cite{fshe03} for 
details of how such fluxes are measured in ISOPHOT maps). 
Together with the flux of 1.58\,Jy at 1.2\,mm given above we can
construct an initial spectral energy distribution, which will allow us a 
comparison with the SEDs of other known Class\,0 sources (especially 
HH\,211-mm) and will give us hints about the nature of this object.
By fitting a modified black body to these measurements we derive 
a bolometric temperature as defined by \cite{ml93} 
of T$_{bol}$\,=\,43\,K, a luminosity of L$_{bol}$\,=\,8\,L$_\odot$, 
a power-law exponent of the frequency dependence of the dust opacity 
of $\beta$\,=\,0.7, and a L$_{smm}$/L$_{bol}$\,=\,0.026. This ratio 
would classify the source as a Class\,0 object, if an extended 
protostellar envelope is found (see \citet{awb00}). Taking the 
evolutionary model of \cite{fshe03} we can further derive an age 
estimate of 29000\,yr, a current stellar mass of 0.1\,M$_\odot$, a 
final stellar mass of 0.5\,M$_\odot$, and an envelope mass of 
1.1\,M$_\odot$ for IC\,348\,MMS. These calculations assume a dust 
opacity of 4\,cm$^2$g$^{-1}$ at 12\,$\mu$m and follow the approach of
\cite{macs98}. The 
observed 1.2\,mm flux ratio of HH\,211-mm to IC\,348\,MMS of 1.2 is in 
good agreement with the modelled envelope mass ratio of 1.3 (see also 
\cite{fshe03}). Thus IC\,348\,MMS seems to be well in the range of 
parameters for a Class\,0 source and may be a bit more evolved than 
HH\,211-mm, although because of the extended 
dust seen in the ISOPHOT maps and its position close to the edge of 
the maps the errors are large. Follow-up with SIRTF is clearly 
necessary to derive the parameters with higher precision.

As a result of our H$_2$ survey, we conclude that at the south-western
edge of IC\,348 star formation is still active. In this region a small
subcluster of embedded young stellar objects is currently
forming. No signs of outflow activity or equally young stars are found
in other parts of IC\,348. At least some of the subcluster members
(HH\,211-mm, IC\,348\,MMS, IC\,348\,IR) are in such early stages of
their evolution (the Class\,0 and Class\,1 phases) that they are
probably much younger than the youngest of the optically visible
objects investigated by \citet{h98}. The unusually large age-spread of
the IC\,348 cluster members that Herbig found may then be understood in
terms of a formation scenario of this cluster, in which subclusters
similar to the one seen here have formed at various times (and
possibly various places) within the cluster. Since the crossing time in  
IC\,348 is on the order of 0.5\,Myr, older subgroups would, however, 
have been spread all over the cluster in the meantime, and should no 
longer appear as localized clumps of similar age. The exception appears 
to be the subcluster discussed in this paper, which is so young that 
its members have not yet moved far from their birthplaces. 

\section{Conclusions}
We have carried out a wide-field H$_2$ survey of the embedded young open 
cluster IC\,348. Outflow activity was only found at its south-western 
limit, in the region of HH\,211. Here, a small subcluster of very deeply
embedded obejcts still in the very early phases of their evolution 
seems to be forming. The considerable age-spread in IC\,348 observed 
by \citet{h98} could be understood if IC\,348 had been 
built up from such subclusters that formed at different times. 

In this part of IC\,348, we observe a large north-south oriented
outflow, and identify a newly discovered mm-object IC\,348\,MMS as its
source. For several other groups of H$_2$ knots, on the other hand, we
cannot identify their sources. 

HH\,211-mm is resolved as a double point-like source in our mm-map, 
although it is not yet clear if the ``secondary'' is a physical 
companion or a dust condensation in the outflow cavity wall. Both 
HH\,211-mm and IC\,348\,MMS are also detected at 160\,$\mu$m and 
200\,$\mu$m with ISOPHOT. Both sources probably are going through 
the Class\,0 phase of their evolution. 

Our deep images in the 1--0 S(1) line of molecular hydrogen trace the
HH\,211 jet and counterjet as highly-collimated straight chains of
knots. Both resemble the molecular jets observed in interferometric
maps in CO and SiO. This jet system appears to be rotated about
3$\degr$  counter-clockwise with respect to the system of the prominent
H$_2$ bow shocks.  We interpret the CO emission ahead of the
south-eastern bow observed by \citet{gg99} -- and difficult to
understand if it was associated with the HH\,211 outflow -- as 
belonging to the IC\,348\,MMS outflow, which is crossing the 
HH\,211 flow in projection in that region.


\begin{acknowledgements}
We thank Manfred Stickel for his help with the data reduction of the 
ISOPHOT data, and Pavel Kroupa for an interesting discussion on 
sequential cluster formation. An anonymous referee has helped to 
clarify this paper. Jochen Eisl\"offel and Dirk Froebrich 
received financial support from DLR through Verbundforschung grant 
50\,OR\,9904\,9, Mark J. McCaughrean through grant 50\,OR\,0004. 
The ISOPHOT data presented in this paper were reduced using PIA, 
which is a joint development by the ESA Astrophysics Division and 
the ISOPHOT Consortium with the collaboration of the Infrared 
Processing and Analysis Center (IPAC). Contributing ISOPHOT Consortium 
institutes are DIAS, RAL, AIP, MPIK and MPIA.
\end{acknowledgements}


\begin{thebibliography}{}


\bibitem[Andr\'e et al. (2000)]{awb00}
Andr\'e, P., Ward-Thompson, D., Barsony, M. 2000, 
Protostars and Planets IV, 
V. Mannings, A.P. Boss, S.S. Russell (eds.) p.59

\bibitem[Avila et al.(2001)]{arc01}
Avila, R., Rodr\'{\i}guez, L.F., Curiel, S. 2001, Revista Mexicana de
Astronom\'{\i}a y Astrof\'{\i}sica, 37, 201

\bibitem[Bachiller, Guilloteau \& Kahane (1987)]{bgk87}
Bachiller, R., Guilloteau, S., Kahane, C. 1987, \aap, 173, 324

\bibitem[Bizenberger et al. (1998)]{bmbts98}
Bizenberger, P., McCaughrean, M.J., Birk, C., Thompson, D., Storz,
C. 1998, SPIE, 3354, 825

\bibitem[Boulard et al. (1995)]{bcmnr95}
Boulard, M.-H., Caux, E., Monin, J.-L., Nadeau, D., Rowlands, N. 1995, 
A\&A, 300, 276

\bibitem[Cernicharo et al. (1985)]{cbd85}
Cernicharo, J., Bachiller, R., Duvert, G. 1985, \aap, 149, 273

\bibitem[Chandler and Richer (2000)]{cr00}
Chandler, C.J., Richer, J.S. 2000, \apj, 530, 851

\bibitem[Chandler and Richer (2001)]{cr01}
Chandler, C.J., Richer, J.S. 2001, \apj, 555, 139

\bibitem[Eisl\"offel (2000)]{e00}
Eisl\"offel, J. 2000, \aap, 354, 236

\bibitem[Eisl\"offel and Mundt (1992)]{em92}
Eisl\"offel, J., Mundt, R. 1992, \aap, 263, 292

\bibitem[Eisl\"offel and Mundt (1994)]{em94}
Eisl\"offel, J., Mundt, R. 1992, \aap, 284, 530

\bibitem[Fendt and Zinnecker (1998)]{fz98}
Fendt, C., Zinnecker, H., 1998, \aap, 334, 750

\bibitem[Froebrich et al. (2003)]{fshe03}
Froebrich, D., Smith, M.D., Hodapp, K.-W., Eisl\"offel, J., 2003, submitted

\bibitem[Gueth and Guilloteau (1999)]{gg99}
Gueth, F., Guilloteau, S. 1999, \aap, 343, 571

\bibitem[Herbig (1998)]{h98}
Herbig, G.H. 1998, \aj, 497, 736

\bibitem[Herbst et al. (1993)]{hbbhmmw93}
Herbst, T.M., Birk, C., Beckwith, S.V.W., Hippler, S., Mc\,Caughrean, M.J.,
Mannucci, F., Wolf, J. 1993, Proc. SPIE, 1946, 605

\bibitem[Kreysa et al. (1998)]{kgg98}
Kreysa, E., Gemuend, H.-P., Gromke, J., et al. 1998, Proc. SPIE, 3357, 319

\bibitem[Lada and Lada (1995)]{ll95}
Lada, E.A., Lada, C.J. 1995, \aj, 109, 1682

\bibitem[Luhman et al. (1998)]{lrll98}
Luhman, K.L., Rieke, G.H., Lada, C.J., Lada, E.A. 1998, \apj, 508, 347

\bibitem[Mc\,Caughrean et al. (1994)]{mrz94}
Mc\,Caughrean, M.J., Rayner, J.T., Zinnecker, H. 1994, \aj, 436, L189

\bibitem[Myers and Ladd (1993)]{ml93}
Myers, P.C., Ladd, E.F. 1993, \apj, 413, L47

\bibitem[Myers et al. (1998)]{macs98}
Myers, P.C., Adams, F.C., Chen, H., Schaff, E. 1998, \apj, 492, 703

\bibitem[Micono et al.(1998)]{mdres98}
Micono, M., Davis, C.J., Ray, T.P., Eisl\"offel, J., Shetrone,
M.D. 1998, \apj 494, L227

\bibitem[Najita et al. (2000)]{ntc00}
Najita, J.R., Tiede, G.P., Carr, J.S. 2000, \apj, 541, 977

\bibitem[Preibisch et al. (1996)]{pzh96}
Preibisch, T., Zinnecker, H., Herbig, G.H. 1996, \aap, 310, 456

\bibitem[Preibisch and Zinnecker (2001)]{pz01}
Preibisch, T., Zinnecker, H. 2001, \aj, 122, 866

\bibitem[Rengel et al. (2002)]{rfeh02}
Rengel, M., Froebrich, D., Hodapp, K., Eisl\"offel, J. 2002, in: The Origins of
Stars and Planets: The VLT View, Jo\~ao Alves \& Mark McCaughrean (ed.)

\bibitem[Rengel et al. (2001)]{rfeh01}
Rengel, M., Froebrich, D., Hodapp, K., Eisl\"offel, J. 2001, in: R. E.
Schielicke (Hrsg.): Astron. Ges. Abstract Ser. 18 (2001), 168 

\bibitem[Smith (1991)]{s91}
Smith, M.D. 1991, \mnras, 253, 175

\bibitem[Strom et al. (1974)]{ssc74}
Strom, S.E., Strom, K.M., Carrasco, L. 1974, PASP, 86, 798

\bibitem[Terquem et al. (1999)]{tepn99}
Terquem, C., Eisl\"offel, J., Papaloizou, J.C.B., Nelson, R.P. 1999,
\apj 512, L131

\end{thebibliography}
\end{document}